\documentclass[aps,prb,reprint,superscriptaddress]{revtex4-2}
\usepackage[T1]{fontenc}
\usepackage{times}
\usepackage{float}
\usepackage{graphicx,color}
\usepackage{amsfonts,amsmath,amssymb,amsbsy}
\usepackage[colorlinks=true, linkcolor=blue, citecolor=blue, urlcolor=blue]{hyperref}
\usepackage{setspace}

\def\D{{\mathcal{D}}}

\let\mathbf=\boldsymbol
\def\blue#1{\textcolor{blue}{#1}}
\def\emph#1{\textcolor{magenta}{#1}}
\def\blue#1{\textcolor{black}{#1}}
\def\emph#1{\textcolor{black}{#1}}
\begin{document}
%%%%%%%%%%%%%%%%%%%%%%%%%%%%%%%%%%%%%%%%%%%%%%%%%%%%%%%%%%%%

\title{Diffusion asymmetry of repulsive skyrmions in structured environment}

\author{Xichao Zhang}
%\email[Corresponding author e-mail:~]{eexzhang@ust.hk}
\affiliation{Department of Applied Physics, Waseda University, Okubo, Shinjuku-ku, Tokyo 169-8555, Japan}
\affiliation{Department of Electronic and Computer Engineering, The Hong Kong University of Science and Technology, Clear Water Bay, Kowloon, Hong Kong, China}
\affiliation{IAS Center for Quantum Matter, The Hong Kong University of Science and Technology, Hong Kong, China}

\author{Charles Reichhardt}
%\email[Email:~]{reichhardt@lanl.gov}
\affiliation{Theoretical Division and Center for Nonlinear Studies, Los Alamos National Laboratory, Los Alamos, New Mexico 87545, USA}

\author{Cynthia J. O. Reichhardt}
%\email[Email:~]{cjrx@lanl.gov}
\affiliation{Theoretical Division and Center for Nonlinear Studies, Los Alamos National Laboratory, Los Alamos, New Mexico 87545, USA}

\author{\\ Qiming Shao}
\email[Corresponding author e-mail:~]{eeqshao@ust.hk}
\affiliation{Department of Electronic and Computer Engineering, The Hong Kong University of Science and Technology, Clear Water Bay, Kowloon, Hong Kong, China}
\affiliation{IAS Center for Quantum Matter, The Hong Kong University of Science and Technology, Hong Kong, China}
\affiliation{Department of Physics, The Hong Kong University of Science and Technology, Clear Water Bay, Kowloon, Hong Kong, China}
\affiliation{State Key Laboratory for Displays and Opto-electronics, The Hong Kong University of Science and Technology, Clear Water Bay, Kowloon, Hong Kong, China}

\author{Rui Zhang}
\email[Corresponding author e-mail:~]{ruizhang@ust.hk}
\affiliation{Department of Physics, The Hong Kong University of Science and Technology, Clear Water Bay, Kowloon, Hong Kong, China}
\affiliation{State Key Laboratory for Displays and Opto-electronics, The Hong Kong University of Science and Technology, Clear Water Bay, Kowloon, Hong Kong, China}
\affiliation{Center for AI for Science, The Hong Kong University of Science and Technology, Clear Water Bay, Kowloon, Hong Kong, China}

\author{Yan Zhou}
%\email[Email:~]{zhouyan@cuhk.edu.cn}
\affiliation{Guangdong Basic Research Center of Excellence for Aggregate Science, School of Science and Engineering, The Chinese University of Hong Kong, Shenzhen, Guangdong 518172, China}

\author{Yongbing Xu}
%\email[Email:~]{ybxu@nju.edu.cn}
\affiliation{National Key Laboratory of Spintronics, Nanjing University, Suzhou 215163, China}
\affiliation{School of Physics, Engineering and Technology, University of York, York, YO10 5DD, UK}

\author{Masahito Mochizuki}
\email[Corresponding author e-mail:~]{masa_mochizuki@waseda.jp}
\affiliation{Department of Applied Physics, Waseda University, Okubo, Shinjuku-ku, Tokyo 169-8555, Japan}

%-%-%-%-%-%-%-%-%-%-%-%-%-%-%-%-%-%-%-%-%-%-%-%-%-%-%-%-%-%-%
\begin{abstract}
\noindent
\textbf{Abstract}\par
\noindent
\emph{The diffusion of matter in structured environments can give rise to emergent phenomena that do not occur in unstructured environments. Here, we report the asymmetric diffusion of magnetic skyrmions in structured chambers, where repulsive skyrmion-skyrmion and skyrmion-environment interactions play vital roles in their diffusive behavior. By fabricating an off-center asymmetric gate separating two chambers, skyrmions could demonstrate asymmetric diffusion through the gate, which depends on the gate symmetry, the gate opening width, and the skyrmion density. Although the diffusion is affected by the skyrmion density, the simulation outcomes are generally in line with rate equation solutions assuming time-independent diffusion rates. Diffusive skyrmions within the chamber can transiently form bonds through repulsive skyrmion-skyrmion interactions, leading to emergent rotational dynamics that is unique to interacting skyrmion systems. Our results uncover asymmetric diffusive behavior of skyrmions, offering insights that will appeal to the wide audience interested in magnetism, active matter, and statistical physics.}
\end{abstract}
%-%-%-%-%-%-%-%-%-%-%-%-%-%-%-%-%-%-%-%-%-%-%-%-%-%-%-%-%-%-%

\date{June 27, 2026}

%\preprint{npj Spintronics}
%\keywords{Skyrmion, Diffusion, Brownian Dynamics, Complex Environments, Spintronics}

\maketitle

\clearpage

%-%-%-%-%-%-%-%-%-%-%-%-%-%-%-%-%-%-%-%-%-%-%-%-%-%-%-%-%-%-%
%\section{Introduction}
%\label{se:Introduction}
%-%-%-%-%-%-%-%-%-%-%-%-%-%-%-%-%-%-%-%-%-%-%-%-%-%-%-%-%-%-%

%-%-%-%-%-%-%-%-%-%-%-%-%-%-%-%-%-%-%-%-%-%-%-%-%-%-%-%-%-%-%
\noindent
\textbf{Introduction}
%-%-%-%-%-%-%-%-%-%-%-%-%-%-%-%-%-%-%-%-%-%-%-%-%-%-%-%-%-%-%

\noindent
Diffusion is one of the most important phenomena in nature that can be observed at a wide range of length and timescales~\cite{Philibert_2006,Mehrer_2009}.
It plays a critical role in many key problems in different fields, including physics~\cite{Fahey_1989,Firstenberg_2013,Huber_2015,Masuda_2017,Yang_2024,Becker_2013}, biology~\cite{Chou_2011,Ricciardi_2013}, and economics~\cite{Hall_2004,Kogut_2011}.
For example, the diffusive transport of particles and particle-like objects is a basic topic in both fundamental research and applied engineering, which covers the flowing phenomena of different phases including gas, liquid, solid, and their mixed phases~\cite{Park_2012,Borg_2012}.
Particularly, the asymmetric diffusion is essential for practical applications~\cite{Packard_2005,Siwy_2005,Shaw_2007,Xu_2016,Galajda_JB2007,Galajda_JMO2008}.
\emph{In principle, the asymmetric diffusion in physical particle systems could also offer a promising route toward unconventional artificial-intelligence (AI) hardware by enabling nonlinear, noise-assisted, and geometry-controlled information processing~\cite{AI1,AI2,AI3}.}

Magnetic skyrmions are a type of topological spin textures that can behave like particle-like objects with chiral dynamic nature~\cite{Mochizuki_JPCM2015,Finocchio_JPD2016,Wiesendanger_NATREVMAT2016,Fert_NATREVMAT2017,Wanjun_PHYSREP2017,Zhang_JPCM2020,Gobel_PHYSREP2021,DelValle_2022,Reichhardt_RMP2022,Ohki_JPCM2024}.
They can be created in bulk magnets, magnetic thin films, and multilayers with antisymmetric or competing exchange interactions~\cite{Mochizuki_JPCM2015,Finocchio_JPD2016,Wiesendanger_NATREVMAT2016,Fert_NATREVMAT2017,Wanjun_PHYSREP2017,Zhang_JPCM2020,Gobel_PHYSREP2021,DelValle_2022,Reichhardt_RMP2022,Ohki_JPCM2024}, such as the Dzyaloshinskii-Moriya (DM) interactions~\cite{Roszler_NATURE2006,Muhlbauer_SCIENCE2009,Yu_NATURE2010,Wanjun_SCIENCE2015,Wanjun_NPHYS2017,Litzius_NPHYS2017}.
Skyrmions stabilized by DM interactions have fixed chirality and helicity, which could give rise to unique dynamics induced by external drives, such as spin currents~\cite{Wanjun_SCIENCE2015,Wanjun_NPHYS2017,Litzius_NPHYS2017,Tomasello_SREP2014,Reichhardt_NC2020,Xichao_PRB2022B,Zhang_Laminar2023}.

Hence, \emph{magnetic} skyrmions could demonstrate various promising transport phenomena due to their unique topology, chirality, and rigidity~\cite{Mochizuki_JPCM2015,Finocchio_JPD2016,Wiesendanger_NATREVMAT2016,Fert_NATREVMAT2017,Wanjun_PHYSREP2017,Zhang_JPCM2020,Gobel_PHYSREP2021,DelValle_2022,Reichhardt_RMP2022,Ohki_JPCM2024}.
A prominent example is the skyrmion Hall effect, which leads to accumulations of skyrmions at device edges~\cite{Wanjun_NPHYS2017,Litzius_NPHYS2017,Zhang_Laminar2023}.
In addition, skyrmions may demonstrate exotic dynamic behaviors that cannot be reproduced by common particles. For example, many skyrmions may flow in nanochannels and generate transport phenomena mimicking chiral fluids~\cite{Zhang_Laminar2023,Raab_PRE2024,Zhang_PNAS2025}.

Recent reports suggest that the thermal effects can result in effective diffusion of skyrmions in ultrathin films and layered heterostructures, where skyrmions could show Brownian-type random walk~\cite{Troncoso_AP2014,Schutte_PRB2014,Tretiakov_PRL2016,Miltat_PRB2018,Nozaki_APL2019,Jing_PRB2021,Zhou_PRB2021,Suzuki_PLA2021,Ishikawa_APL2021,Weissenhofer_NJP2020,Weissenhofer_PRL2021,Weissenhofer_PRB2023,Gruber_AM2023,Dohi_NC2023,Kerber_PRA2021} and may demonstrate directional diffusion in the presence of a thermal gradient~\cite{Kong_PRL2013,Lin_PRL2014,Reichhardt_NJP2016,Reichhardt_JPCM2019,Wang_NE2020,Kong_PRB2021}.
The diffusion of an individual skyrmion can also become anisotropic by applying an in-plane magnetic field to break the rotational symmetry of the skyrmion~\cite{Kerber_PRA2021}.

More intriguingly, recent investigations have uncovered a topology-dependent Brownian gyromotion of skyrmions, which is an emerging phenomenon that becomes pronounced when diffusive skyrmions interact with confinement~\cite{Zhao_PRL2020,Miki_JPSJ2021,Zhang_NL2023}.
The Brownian gyromotion may lead to topological sorting phenomena when the confinement environment is also of chiral nature~\cite{Zhang_NL2023}.
However, many diffusive properties of \emph{magnetic} skyrmions, especially in \emph{structured environments}, still remain elusive and largely unexplored.
Understanding the diffusion of skyrmions is essentially important for their functionalization~\cite{Nozaki_APL2019,Jakobs_NN2019,Pinna_PRA2018,Zhu_PRA2020,Jibiki_APL2020,Miki_JSAP2025,Suzuki_2025,Winkler_2025} as well as their interdisciplinary development~\cite{Silva_PRL2025,Zhang_P2025}.

%%%%%%%%%%%%%%%%%%%%%%%%%%%%%%%%%%%%%%%%%%%%%%%%%%%%%%%%%%%%
\begin{figure*}[t]
\centerline{\includegraphics[width=0.63\textwidth]{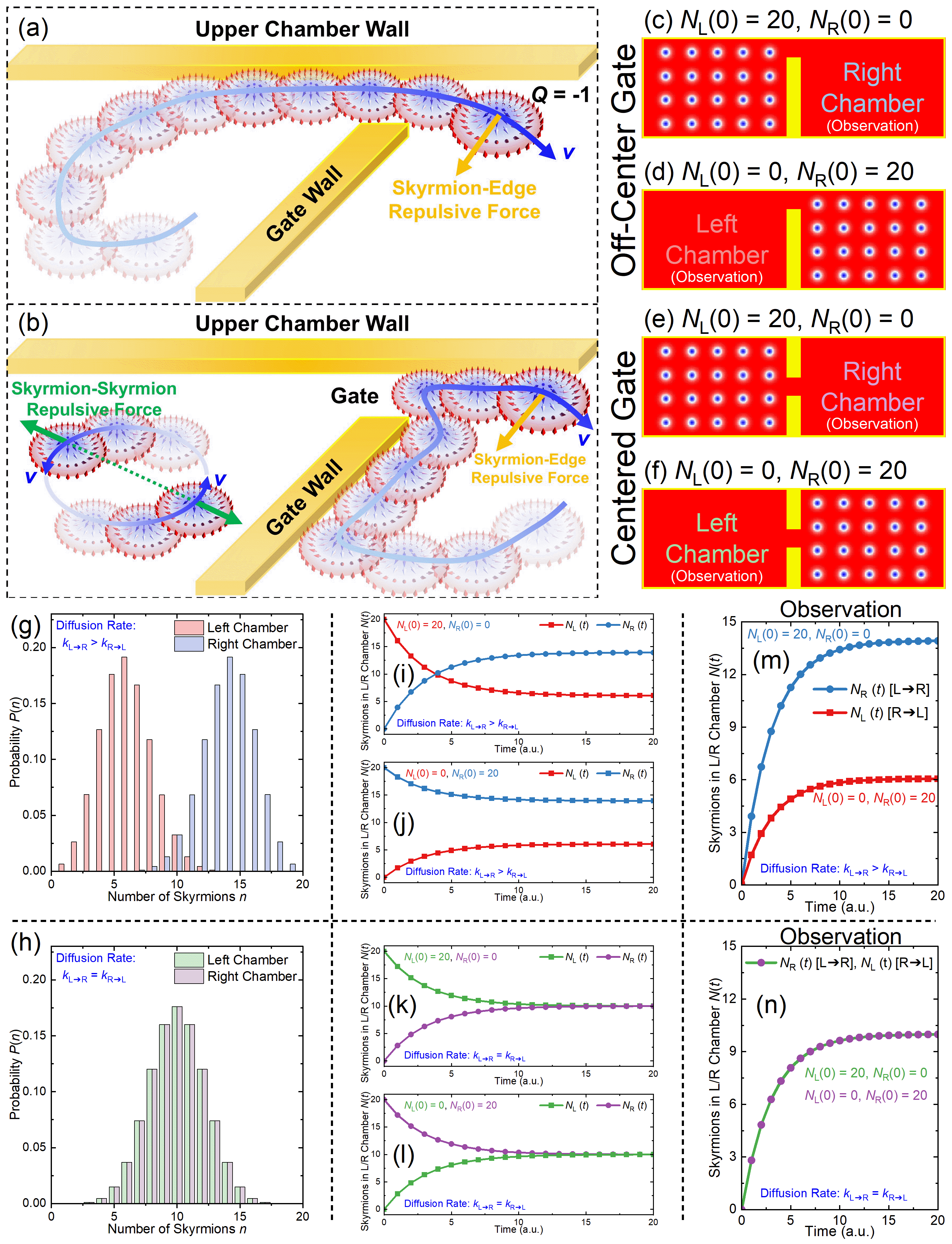}}
\caption{%
\textbf{Theoretical probability distribution and diffusion dynamics.}
(a) Schematic illustration showing that a diffusive skyrmion with $Q=-1$ in the left chamber can pass through the \emph{off-center asymmetric gate (OAG)} due to its interaction with the upper chamber wall.
(b) Schematic illustration showing that a diffusive skyrmion with $Q=-1$ in the right chamber is unable to pass through the \emph{OAG} due to its interaction with the upper chamber wall. Two diffusive skyrmions in the left chamber demonstrate counterclockwise rotation when interacting with each other, forming a transient binary skyrmion system.
(c) Top view of the simulated \emph{OAG} system. The out-of-plane spin component is color coded: blue is into the plane, red is out of the plane, white is in-plane. At $t=0$ ns, $20$ skyrmions with $Q=-1$ are relaxed in the left chamber [$N_{\text{L}}(0)=20$] and no skyrmions are placed in the right chamber [$N_{\text{R}}(0)=0$].
The chamber walls are indicated by yellow areas.
(d) The \emph{OAG} system with $N_{\text{L}}(0)=0$ and $N_{\text{R}}(0)=20$.
(e) The \emph{centered symmetric-gate (CSG)} system with $N_{\text{L}}(0)=20$ and $N_{\text{R}}(0)=0$.
(f) The \emph{CSG} system with $N_{\text{L}}(0)=0$ and $N_{\text{R}}(0)=20$.
(g) Typical probability distribution of the \emph{OAG} system. In principle, skyrmions showing clockwise Brownian gyromotion ($Q=-1$) is easier to pass through the \emph{OAG} from the left side. Hence, here we assume that the diffusion rate $k_{\text{L}\rightarrow\text{R}}=0.23$ and $k_{\text{R}\rightarrow\text{L}}=0.1$.
The subscript ``$\text{L}\rightarrow\text{R}$'' denotes skyrmion diffusion from the left chamber to the right, while ``$\text{R}\rightarrow\text{L}$'' denotes the diffusion in the opposite direction.
(h) Typical probability distribution of the \emph{CSG} system with $k_{\text{L}\rightarrow\text{R}}=k_{\text{R}\rightarrow\text{L}}=0.165$.
(i) Time-dependent numbers of skyrmions in the left [$N_{\text{L}}(t)$] and right [$N_{\text{R}}(t)$] chambers of the \emph{OAG} system, where $N_{\text{L}}(0)=20$ and $N_{\text{R}}(0)=0$.
(j) $N_{\text{L}}(t)$ and $N_{\text{R}}(t)$ of the \emph{OAG} system, where $N_{\text{L}}(0)=0$ and $N_{\text{R}}(0)=20$.
(k) $N_{\text{L}}(t)$ and $N_{\text{R}}(t)$ of the \emph{CSG} system, where $N_{\text{L}}(0)=20$ and $N_{\text{R}}(0)=0$.
(l) $N_{\text{L}}(t)$ and $N_{\text{R}}(t)$ of the \emph{CSG} system, where $N_{\text{L}}(0)=0$ and $N_{\text{R}}(0)=20$.
(m) Time-dependent $N_{\text{R} (\text{L}\rightarrow\text{R})}(t)$ and $N_{\text{L} (\text{R}\rightarrow\text{L})}(t)$ of the \emph{OAG} system.
(n) $N_{\text{R} (\text{L}\rightarrow\text{R})}(t)$ and $N_{\text{L} (\text{R}\rightarrow\text{L})}(t)$ of the \emph{CSG} system.
}
\label{FIG1}
\end{figure*}
%%%%%%%%%%%%%%%%%%%%%%%%%%%%%%%%%%%%%%%%%%%%%%%%%%%%%%%%%%%%

%%%%%%%%%%%%%%%%%%%%%%%%%%%%%%%%%%%%%%%%%%%%%%%%%%%%%%%%%%%%
\begin{figure*}[t]
\centerline{\includegraphics[width=0.90\textwidth]{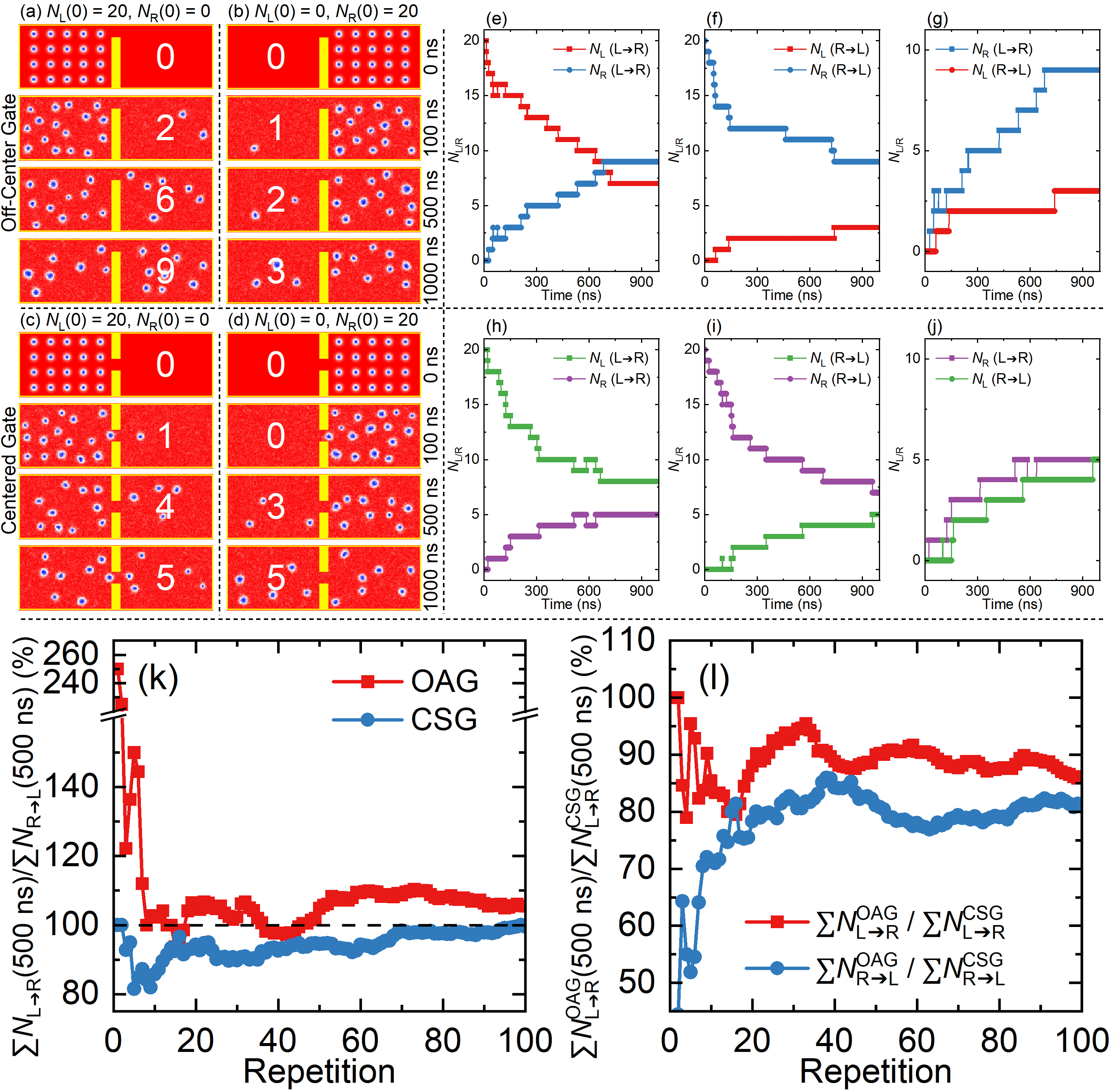}}
\caption{%
\textbf{Simulated skyrmion diffusion in the \emph{symmetric and asymmetric} systems.}
(a) Selected snapshots of the \emph{OAG} system with $N_{\text{L}}(0)=20$ and $N_{\text{R}}(0)=0$.
(b) Selected snapshots of the \emph{OAG} system with $N_{\text{L}}(0)=0$ and $N_{\text{R}}(0)=20$.
(c) Selected snapshots of the \emph{CSG} system with $N_{\text{L}}(0)=20$ and $N_{\text{R}}(0)=0$.
(d) Selected snapshots of the \emph{CSG} system with $N_{\text{L}}(0)=0$ and $N_{\text{R}}(0)=20$.
Simulations in (a-d) were performed with the parameters: $W=36$ nm, $T=150$ K, and $S=304$.
$N_{\text{L}}(t)$ and $N_{\text{R}}(t)$ in the simulated \emph{OAG} system are given for (e) diffusion in the $\text{L}\rightarrow\text{R}$ direction and (f) diffusion in the $\text{R}\rightarrow\text{L}$ direction, respectively. Time-dependent $N_{\text{R} (\text{L}\rightarrow\text{R})}$ and $N_{\text{L} (\text{R}\rightarrow\text{L})}$ are given in (g) to show the diffusion asymmetry.
$N_{\text{L}}(t)$ and $N_{\text{R}}(t)$ in the simulated \emph{CSG} system are given for (h) diffusion in the $\text{L}\rightarrow\text{R}$ direction and (i) diffusion in the $\text{R}\rightarrow\text{L}$ direction, respectively. Time-dependent $N_{\text{R} (\text{L}\rightarrow\text{R})}$ and $N_{\text{L} (\text{R}\rightarrow\text{L})}$ are given in (j) to show the diffusion symmetry.
(k) The repetition-dependent ratio between $\sum_{S} N_{\text{R} (\text{L}\rightarrow\text{R})}$ and $\sum_{S} N_{\text{L} (\text{R}\rightarrow\text{L})}$ is obtained based on the simulated outcome at $t=500$ ns for $100$ repetitions of the same simulation but with different thermal seeds $S$. Other parameters are fixed: $W=36$ nm and $T=150$ K.
(l) The repetition-dependent ratio between $\sum_{S} N_{\text{R} (\text{L}\rightarrow\text{R})}^{\text{OAG}}$ and $\sum_{S} N_{\text{R} (\text{L}\rightarrow\text{R})}^{\text{CSG}}$, and the repetition-dependent ratio between $\sum_{S} N_{\text{L} (\text{R}\rightarrow\text{L})}^{\text{OAG}}$ and $\sum_{S} N_{\text{L} (\text{R}\rightarrow\text{L})}^{\text{CSG}}$ are obtained based on the simulated outcome at $t=500$ ns for $100$ repetitions of the same simulation but with different thermal seeds $S$.
}
\label{FIG2}
\end{figure*}
%%%%%%%%%%%%%%%%%%%%%%%%%%%%%%%%%%%%%%%%%%%%%%%%%%%%%%%%%%%%

%%%%%%%%%%%%%%%%%%%%%%%%%%%%%%%%%%%%%%%%%%%%%%%%%%%%%%%%%%%%
\begin{figure}[t]
\centerline{\includegraphics[width=0.49\textwidth]{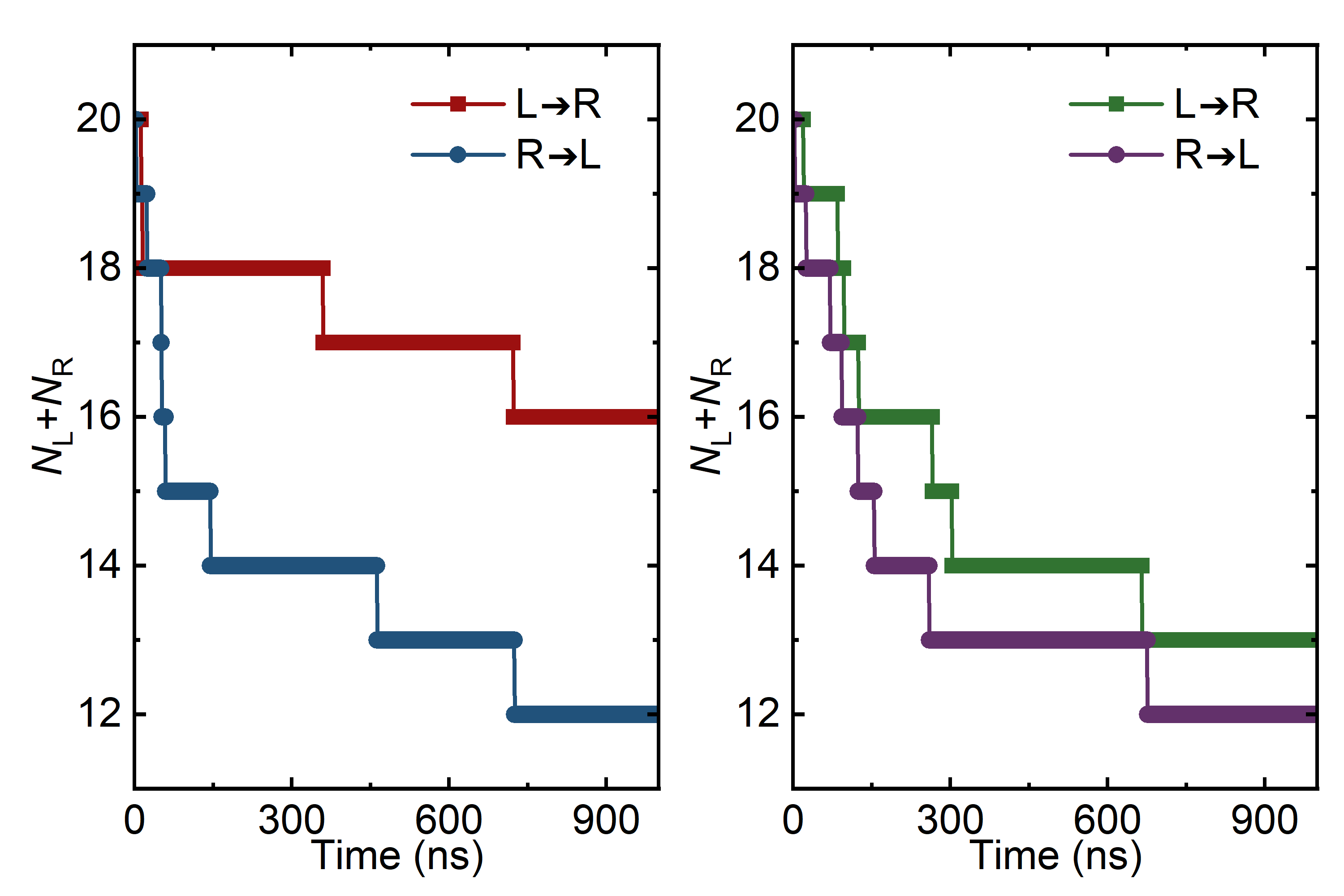}}
\caption{%
\emph{\textbf{Thermal annihilation of skyrmions during their diffusion.}
(a) Time-dependent total number of skyrmions $N_{\text{L}}(t)+N_{\text{R}}(t)$, including skyrmions in the left chamber $N_{\text{L}}(t)$ and skyrmions in the right chamber $N_{\text{R}}(t)$, for the system with an asymmetric gate.
``L$\rightarrow$ R'' indicates the case where skyrmions diffuse from the left chamber to the right chamber with $N_{\text{L}}(0)=20$ and $N_{\text{R}}(0)=0$.
``R$\rightarrow$ L'' indicates the case where skyrmions diffuse from the right chamber to the left chamber with $N_{\text{L}}(0)=0$ and $N_{\text{R}}(0)=20$.
(b) Time-dependent total number of skyrmions $N_{\text{L}}(t)+N_{\text{L}}(t)$ for the system with a symmetric gate.
The above four simulations were performed with the same parameters: $W=36$ nm, $T=150$ K, and $S=304$. More details are given in the caption of Fig.~\ref{FIG2}.
The simulation results show that certain number of skyrmions are annihilated for both the symmetric and asymmetric systems during the diffusion process due to the thermal fluctuations.}
}
\label{FIG3}
\end{figure}
%%%%%%%%%%%%%%%%%%%%%%%%%%%%%%%%%%%%%%%%%%%%%%%%%%%%%%%%%%%%

%%%%%%%%%%%%%%%%%%%%%%%%%%%%%%%%%%%%%%%%%%%%%%%%%%%%%%%%%%%%
\begin{figure*}[t]
\centerline{\includegraphics[width=0.98\textwidth]{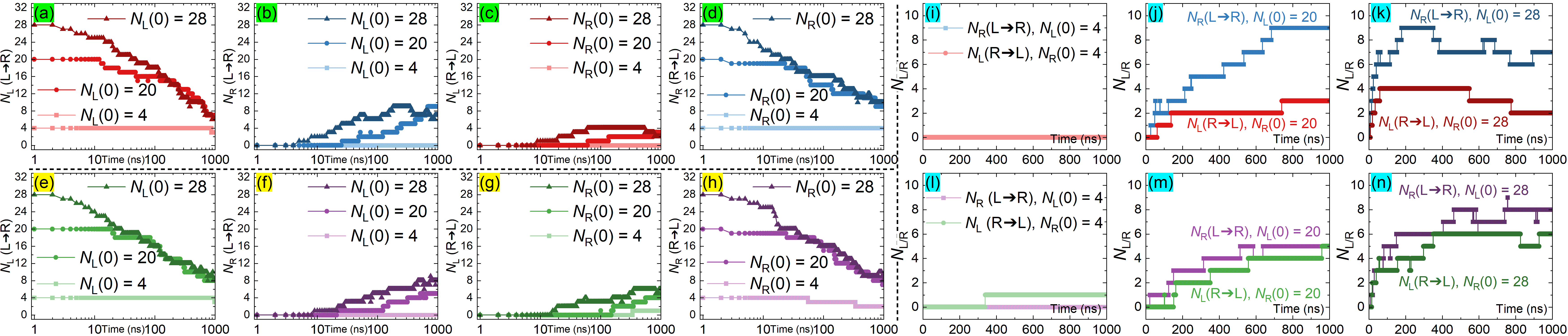}}
\caption{%
\emph{\textbf{Influence of the initial skyrmion density on the skyrmion diffusion.}
(a) $N_{\text{L} (\text{L}\rightarrow\text{R})}(t)$ for simulated \emph{OAG} systems with different values of $N_{\text{L}}(0)$.
(b) $N_{\text{R} (\text{L}\rightarrow\text{R})}(t)$ for simulated \emph{OAG} systems with different values of $N_{\text{L}}(0)$.
(c) $N_{\text{L} (\text{R}\rightarrow\text{L})}(t)$ for simulated \emph{OAG} systems with different values of $N_{\text{R}}(0)$.
(d) $N_{\text{R} (\text{R}\rightarrow\text{L})}(t)$ for simulated \emph{OAG} systems with different values of $N_{\text{R}}(0)$.
(e) $N_{\text{L} (\text{L}\rightarrow\text{R})}(t)$ for simulated \emph{CSG} systems with different values of $N_{\text{L}}(0)$.
(f) $N_{\text{R} (\text{L}\rightarrow\text{R})}(t)$ for simulated \emph{CSG} systems with different values of $N_{\text{L}}(0)$.
(g) $N_{\text{L} (\text{R}\rightarrow\text{L})}(t)$ for simulated \emph{CSG} systems with different values of $N_{\text{R}}(0)$.
(h) $N_{\text{R} (\text{R}\rightarrow\text{L})}(t)$ for simulated \emph{CSG} systems with different values of $N_{\text{R}}(0)$.
A comparison between $N_{\text{R} (\text{L}\rightarrow\text{R})}(t)$ and $N_{\text{L} (\text{R}\rightarrow\text{L})}(t)$ are given respectively for \emph{OAG} systems with (i) $N_{\text{L}/\text{R}}(0)=4$, (j) $N_{\text{L}/\text{R}}(0)=20$, and (k) $N_{\text{L}/\text{R}}(0)=28$.
A comparison between $N_{\text{R} (\text{L}\rightarrow\text{R})}(t)$ and $N_{\text{L} (\text{R}\rightarrow\text{L})}(t)$ are given respectively for \emph{CSG} systems with (l) $N_{\text{L}/\text{R}}(0)=4$, (m) $N_{\text{L}/\text{R}}(0)=20$, and (n) $N_{\text{L}/\text{R}}(0)=28$.
Here, all these simulations were performed with the same parameters as in Fig.~\ref{FIG2}.}
}
\label{FIG4}
\end{figure*}
%%%%%%%%%%%%%%%%%%%%%%%%%%%%%%%%%%%%%%%%%%%%%%%%%%%%%%%%%%%%

%%%%%%%%%%%%%%%%%%%%%%%%%%%%%%%%%%%%%%%%%%%%%%%%%%%%%%%%%%%%
\begin{figure}[t]
\centerline{\includegraphics[width=0.49\textwidth]{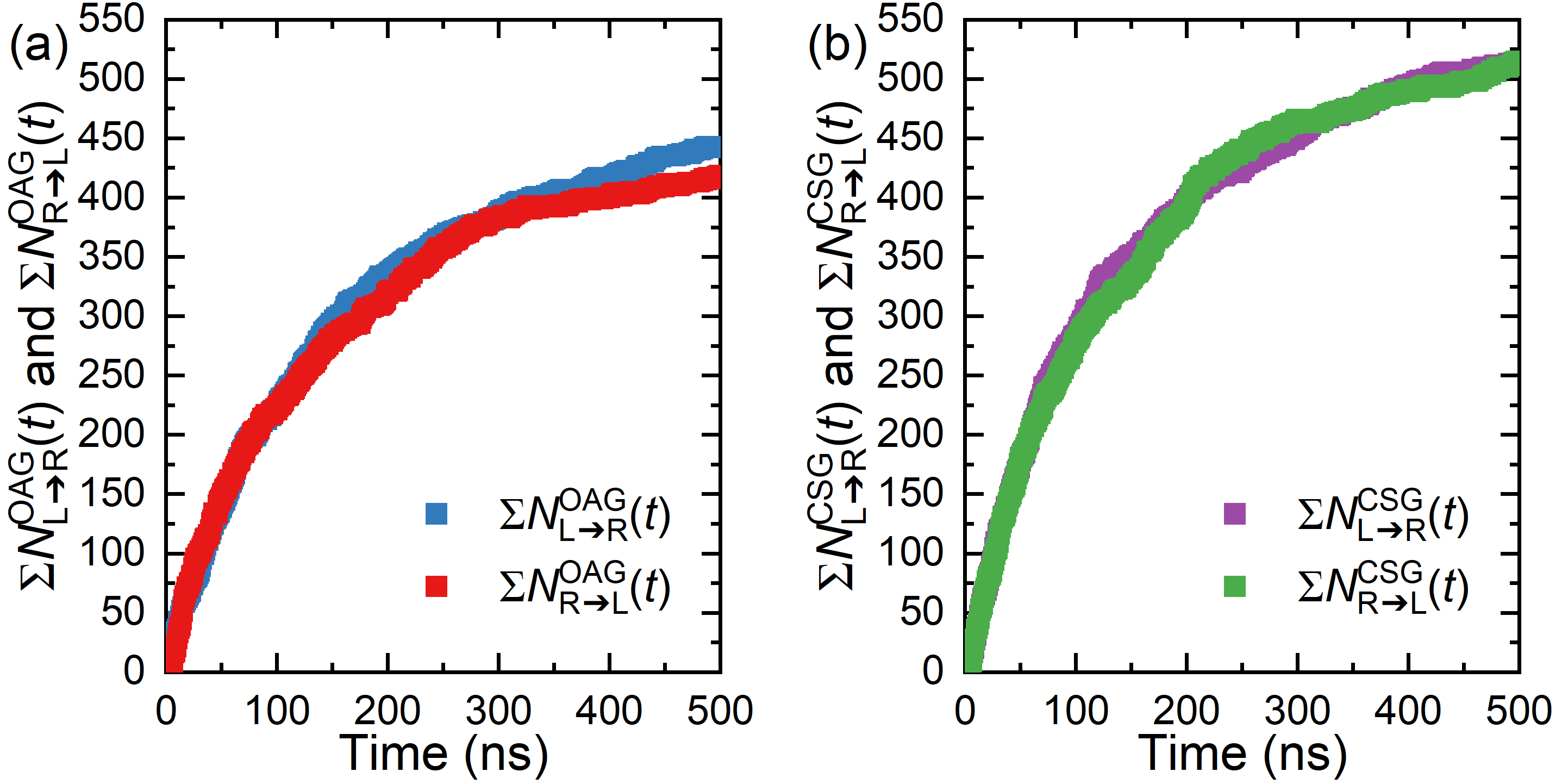}}
\caption{%
\emph{\textbf{Summation of the time-dependent number of skyrmions found in the observation chamber for $100$ repetitions of simulations with different random seeds.}
(a) $\sum_{S}N_{\text{R} (\text{L}\rightarrow\text{R})}^{\text{OAG}}(t)$ and $\sum_{S}N_{\text{L} (\text{R}\rightarrow\text{L})}^{\text{OAG}}(t)$ for $t=0$-$500$ ns.
(b) $\sum_{S}N_{\text{R} (\text{L}\rightarrow\text{R})}^{\text{CSG}}(t)$ and $\sum_{S}N_{\text{L} (\text{R}\rightarrow\text{L})}^{\text{CSG}}(t)$ for $t=0$-$500$ ns.
All these simulations were performed with parameters: $T=150$ K and $W=36$ nm.
As the diffusion rates in simulation experiments depend on several factors and vary with time, here we focus on the diffusion behavior happened during the early stages of the simulation (i.e., $t=0$-$500$ ns), where the change of diffusion rates could be small. The diffusion asymmetry could be found in the OAG system, while the diffusion is generally symmetric in the CSG system.}
}
\label{FIG5}
\end{figure}
%%%%%%%%%%%%%%%%%%%%%%%%%%%%%%%%%%%%%%%%%%%%%%%%%%%%%%%%%%%%

%%%%%%%%%%%%%%%%%%%%%%%%%%%%%%%%%%%%%%%%%%%%%%%%%%%%%%%%%%%%
\begin{figure*}[t]
\centerline{\includegraphics[width=0.65\textwidth]{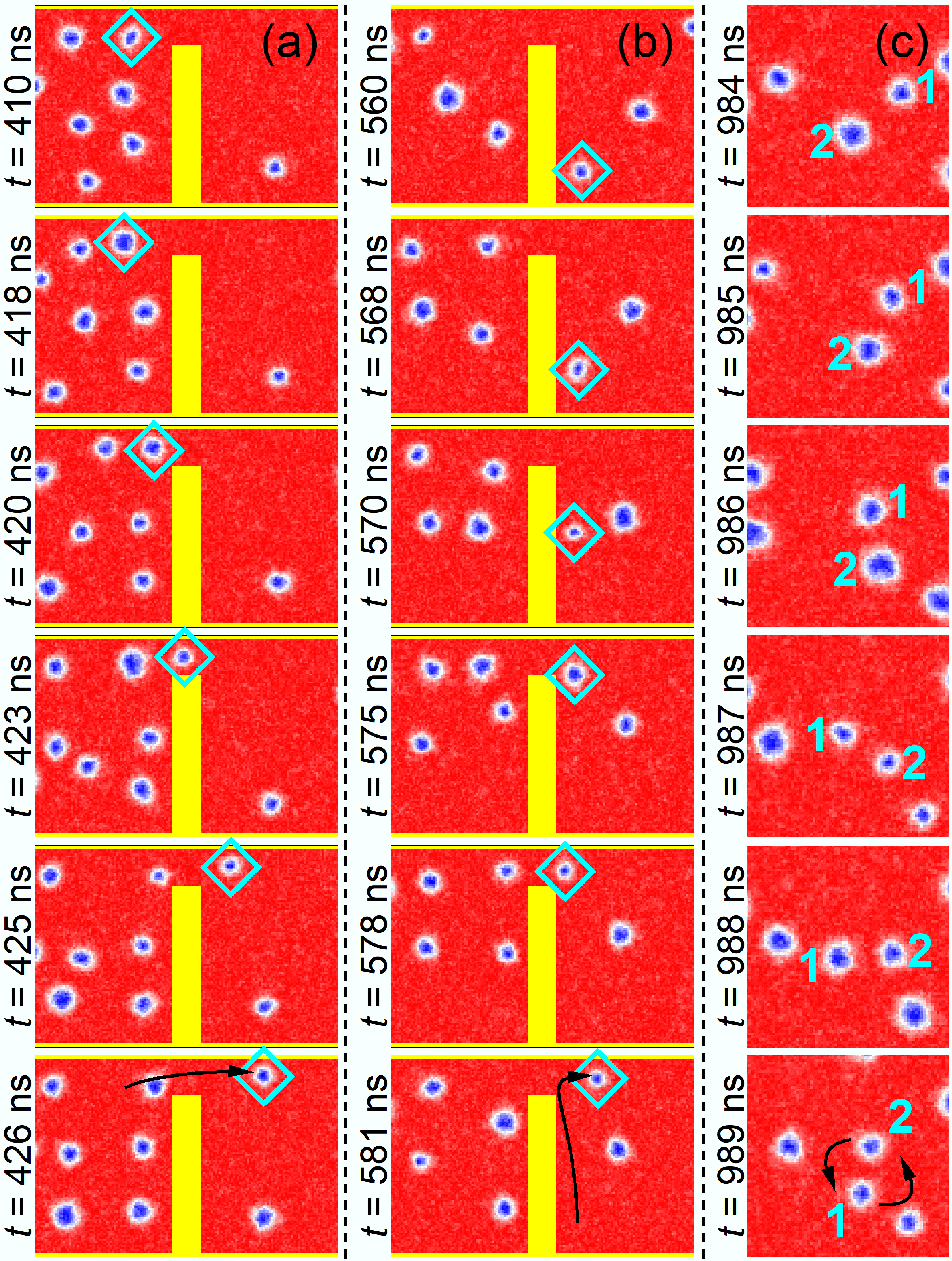}}
\caption{%
\textbf{A diffusive skyrmion interacting with the \emph{OAG} geometry and two diffusive skyrmions inside the chamber interacting with each other.}
(a) Selected close-up snapshots showing that a diffusive skyrmion with $Q=-1$ first approaches with the upper chamber wall and then passes through the \emph{OAG}. The skyrmion is marked by a cyan diamond.
(b) Selected close-up snapshots showing that a diffusive skyrmion with $Q=-1$ first approaches the gate wall and then moves toward the upper chamber wall. The skyrmion does not pass through the \emph{OAG} due to its clockwise rotation sense when interacting with \emph{the upper chamber wall}.
(c) Selected close-up snapshots showing that two diffusive skyrmions with $Q=-1$ inside the right chamber of the device can interact with each other and demonstrate counterclockwise rotation, which leads to a transient binary skyrmion system, \emph{although the interaction between them is of repulsive nature.}
All snapshots are selected from the simulation for Fig.~\ref{FIG2}(a).
}
\label{FIG6}
\end{figure*}
%%%%%%%%%%%%%%%%%%%%%%%%%%%%%%%%%%%%%%%%%%%%%%%%%%%%%%%%%%%%

%%%%%%%%%%%%%%%%%%%%%%%%%%%%%%%%%%%%%%%%%%%%%%%%%%%%%%%%%%%%
\begin{figure}[t]
\centerline{\includegraphics[width=0.49\textwidth]{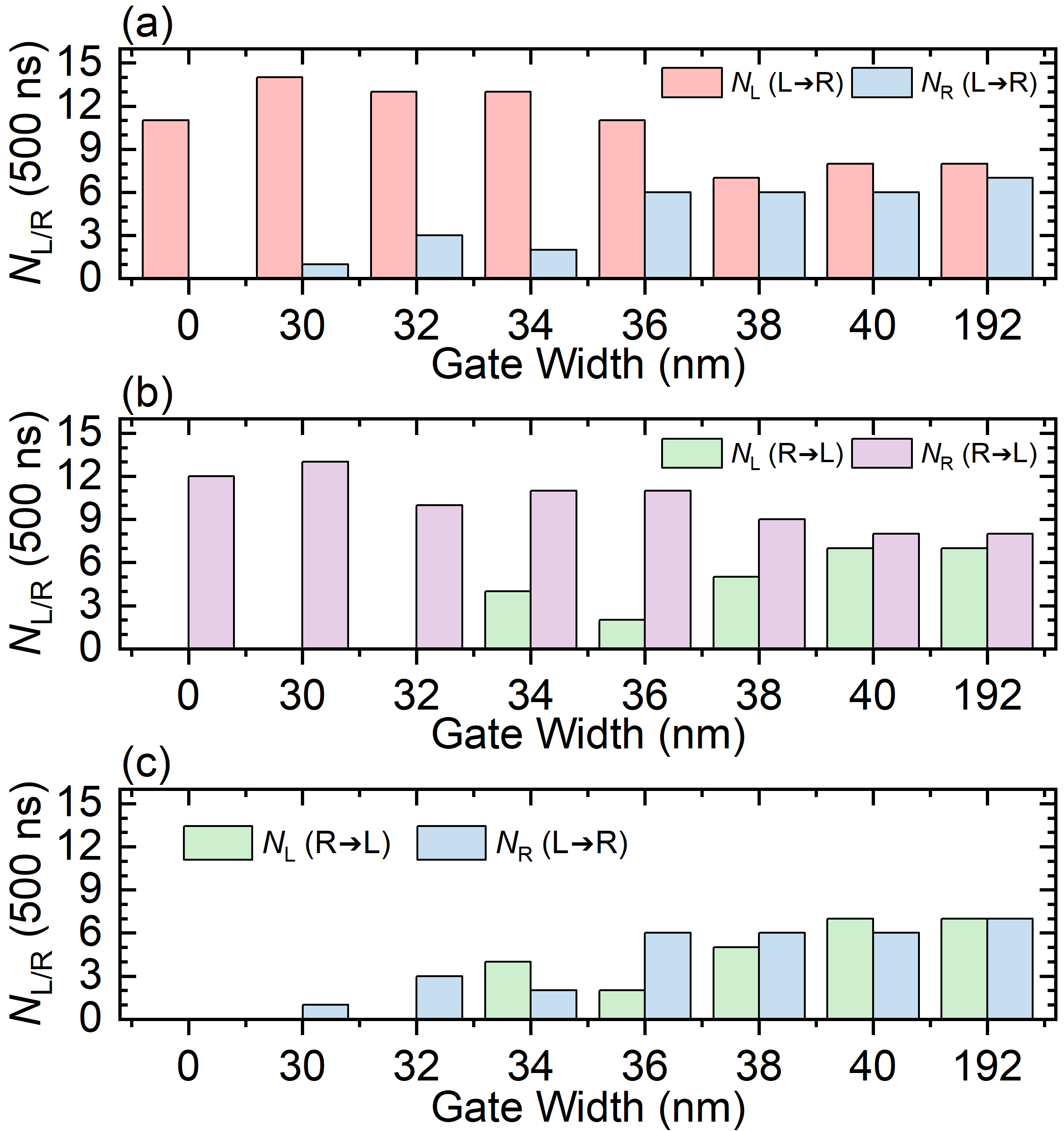}}
\caption{%
\emph{\textbf{Effect of OAG opening width on the simulated outcomes.}
(a) $N_{\text{L} (\text{L}\rightarrow\text{R})}$ and $N_{\text{R} (\text{L}\rightarrow\text{R})}$ obtained at $t=500$ ns for different asymmetric gate width, where $N_{\text{L}}(0)=20$ and $N_{\text{R}}(0)=0$.
(b) $N_{\text{L} (\text{R}\rightarrow\text{L})}$ and $N_{\text{R} (\text{R}\rightarrow\text{L})}$ obtained at $t=500$ ns for different asymmetric gate width, where $N_{\text{L}}(0)=0$ and $N_{\text{R}}(0)=20$.
(c) Comparison between $N_{\text{R} (\text{L}\rightarrow\text{R})}$ and $N_{\text{L} (\text{R}\rightarrow\text{L})}$ obtained at $t=500$ ns.
Here, all these simulations were performed with the same parameters as in Fig.~\ref{FIG2}.}
}
\label{FIG7}
\end{figure}
%%%%%%%%%%%%%%%%%%%%%%%%%%%%%%%%%%%%%%%%%%%%%%%%%%%%%%%%%%%%

%%%%%%%%%%%%%%%%%%%%%%%%%%%%%%%%%%%%%%%%%%%%%%%%%%%%%%%%%%%%
\begin{figure}[t]
\centerline{\includegraphics[width=0.45\textwidth]{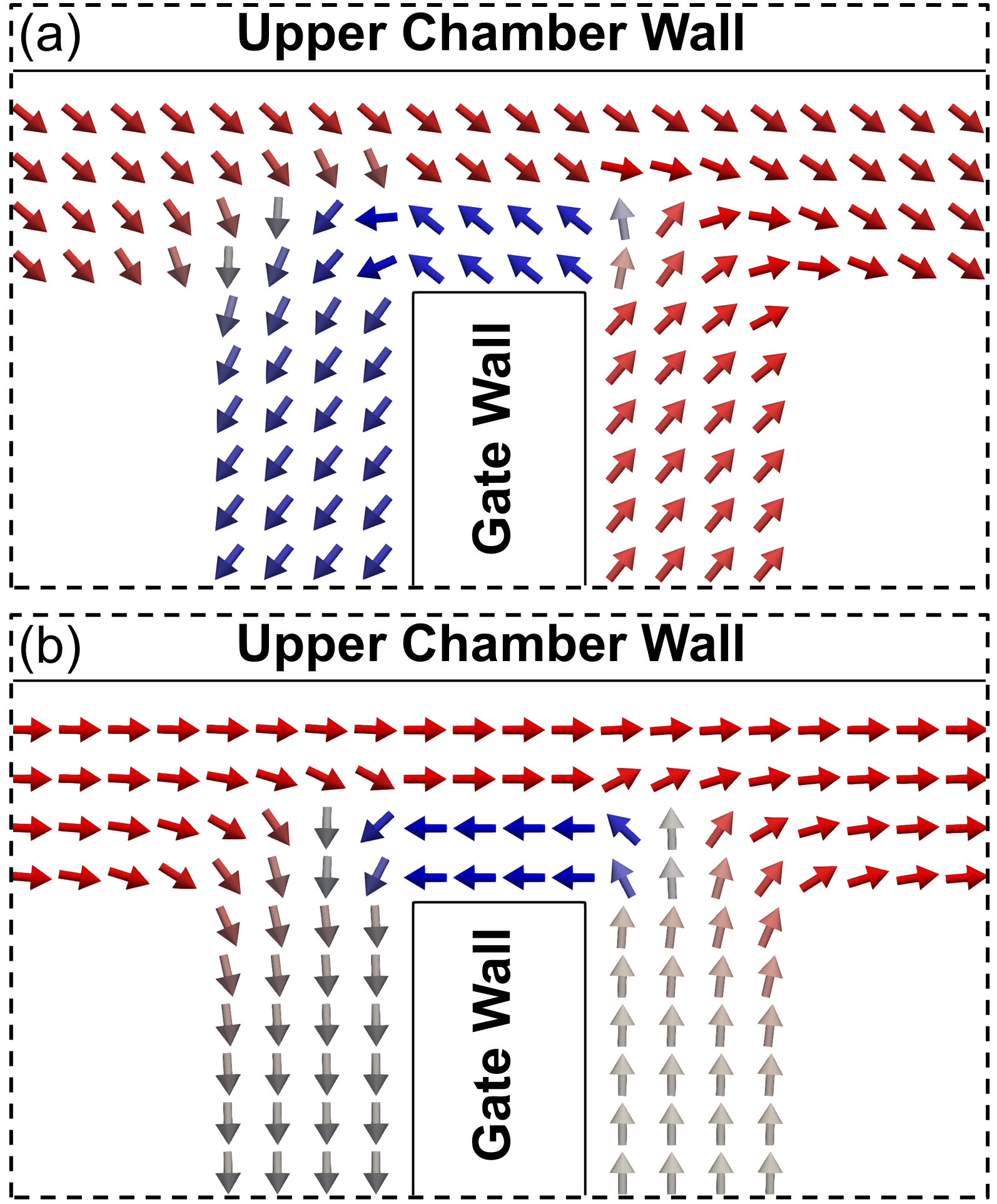}}
\caption{%
\blue{\textbf{Velocity field of skyrmions moving near the upper chamber wall and gate wall obtained based on the Thiele model for $Q=-1$.}
(a) We assumed that $\left|\alpha\D\right|=10$, i.e., the damping $\alpha$ is nonzero.
(b) We assumed that $\left|\alpha\D\right|=0$, i.e., the damping $\alpha$ equals zero.
Other parameters are given according to the OAG system investigated in Fig.~\ref{FIG2}(c).}
}
\label{FIG8}
\end{figure}
%%%%%%%%%%%%%%%%%%%%%%%%%%%%%%%%%%%%%%%%%%%%%%%%%%%%%%%%%%%%

%-%-%-%-%-%-%-%-%-%-%-%-%-%-%-%-%-%-%-%-%-%-%-%-%-%-%-%-%-%-%
%\section{Theoretical Analysis}
%\label{se:Theory}
%-%-%-%-%-%-%-%-%-%-%-%-%-%-%-%-%-%-%-%-%-%-%-%-%-%-%-%-%-%-%

%-%-%-%-%-%-%-%-%-%-%-%-%-%-%-%-%-%-%-%-%-%-%-%-%-%-%-%-%-%-%
%\vbox{}
%\noindent
%\textbf{Results and Discussion}
%-%-%-%-%-%-%-%-%-%-%-%-%-%-%-%-%-%-%-%-%-%-%-%-%-%-%-%-%-%-%

%-%-%-%-%-%-%-%-%-%-%-%-%-%-%-%-%-%-%-%-%-%-%-%-%-%-%-%-%-%-%
\vbox{}
\noindent
\textbf{Results}
%-%-%-%-%-%-%-%-%-%-%-%-%-%-%-%-%-%-%-%-%-%-%-%-%-%-%-%-%-%-%

%-%-%-%-%-%-%-%-%-%-%-%-%-%-%-%-%-%-%-%-%-%-%-%-%-%-%-%-%-%-%
\noindent
\textbf{Model and theoretical analysis}
%-%-%-%-%-%-%-%-%-%-%-%-%-%-%-%-%-%-%-%-%-%-%-%-%-%-%-%-%-%-%

\noindent
In this work, we report the asymmetric diffusion of many interacting \emph{magnetic} skyrmions at finite temperature but in the absence of thermal gradient. The asymmetry of diffusion is generated by effective but nonsustained interactions between magnetic skyrmions and \emph{an off-center asymmetric gate (OAG)} geometry.
The \emph{repulsive} skyrmion-skyrmion interaction also plays \emph{an important} role in the diffusion process.

In Figs.~\ref{FIG1}(a) and~\ref{FIG1}(b), we first illustrate the interaction between an individual N{\'e}el-type skyrmion and an \emph{OAG} driven by thermal fluctuations, which is essential for the asymmetric diffusion of many skyrmions in the \emph{OAG} chamber [Fig.~\ref{FIG1}(c)].
The dynamics of the skyrmion can be described by the Thiele equation~\cite{Thiele_PRL1973,Tomasello_SREP2014}
$\boldsymbol{G}\times\boldsymbol{v}-\alpha\boldsymbol{\D}\cdot\boldsymbol{v}+\boldsymbol{F}_{\text{w}}+\boldsymbol{F}_{\text{f}}=\boldsymbol{0}$,
where $\boldsymbol{v}$ is the skyrmion velocity, $\boldsymbol{G}=(0,0,-4\pi Q)$ is the gyromagnetic coupling vector associated with the Magnus force $\boldsymbol{F}_{\text{M}}=\boldsymbol{G}\times\boldsymbol{v}$, $\boldsymbol{\D}$ is a dissipative tensor, and $\alpha$ is the damping parameter.
$\boldsymbol{F}_{\text{w}}$ is a repulsive force acting the skyrmion when it interacts with the wall barrier.
$\boldsymbol{F}_{\text{f}}$ is a Gaussian stochastic force describing the thermal effects (see Methods), which leads to the Brownian motion of the skyrmion but has a zero average.
We focus on the skyrmion carrying the topological charge of $Q=\frac{1}{4\pi}\int\boldsymbol{m}\cdot(\frac{\partial\boldsymbol{m}}{\partial x}\times\frac{\partial\boldsymbol{m}}{\partial y})dxdy=-1$.

In principle, $F_{\text{w}}=0$ when the skyrmion is far from the wall and therefore, the skyrmion will demonstrate $\boldsymbol{F}_{\text{f}}$-induced Brownian random walk without gyromotion~\cite{Zhao_PRL2020,Zhang_NL2023}.
However, when the skyrmion is close to the wall during its diffusion, a nonzero repulsive $\boldsymbol{F}_{\text{w}}$ will act on the skyrmion, of which the direction is perpendicular to the wall and the strength $F_{\text{w}}$ significantly increases with decreasing spacing between the skyrmion and the wall.
\blue{As the damping parameter is usually small $\alpha<1$ and the magnitude of the dissipation tensor is also small for nanoscale skyrmions~\cite{Wanjun_NPHYS2017}, the skyrmion will then exhibit Brownian gyromotion guided by the wall for a short transient period as long as the skyrmion-wall interaction persists (i.e., $F_{\text{w}}\neq 0$), which could be basically described by $\boldsymbol{G}\times\boldsymbol{v}=-\boldsymbol{F}_{\text{w}}$ assuming $\left|\alpha\boldsymbol{\D}\cdot\boldsymbol{v}\right|\sim 0$.}
It shows that the direction of $\boldsymbol{v}$ is mainly determined by the topology of the skyrmion $Q$ and the direction of $\boldsymbol{F}_{\text{w}}$.
\blue{However, it should be noted that a nonzero but small $\left|\alpha\boldsymbol{\D}\cdot\boldsymbol{v}\right|$ is essential for the diffusion asymmetry to arise near the OAG, which will be discussed later.}

Consequently, the skyrmion with $Q=-1$ will demonstrate clockwise Brownian gyromotion along the chamber wall or gate wall [Figs.~\ref{FIG1}(a) and~\ref{FIG1}(b)].
Also, the skyrmion may accelerate when it approaches the wall ($dF_{\text{w}}/dt>0$) and decelerate when it detaches from the wall ($dF_{\text{w}}/dt<0$).
Due to the Brownian gyromotion guided by chamber and gate walls, the skyrmion with $Q=-1$ placed in the left chamber could pass through the \emph{OAG} as illustrated in Fig.~\ref{FIG1}(a), while it may not be able to pass through the \emph{OAG} from the right chamber [Fig.~\ref{FIG1}(b)].
\emph{Note that placing the OAG near the lower chamber wall will result in asymmetric diffusion in the opposite direction provided that the topological charge of skyrmions remain unchanged (i.e., $Q=-1$). It is worth mentioning that the internal helicity configuration of the skyrmion should not affect its diffusive behavior, as $Q$ is independent of the helicity structure.}

\emph{We also note that diffusive skyrmions interacting with each other inside the chamber may demonstrate Brownian rotation as the repulsive skyrmion-skyrmion interaction can result in a force $\boldsymbol{F}_{\text{s}}$ acting on the skyrmion, similar to the effect of $\boldsymbol{F}_{\text{w}}$. For example, the two skyrmions with $Q=-1$ in the left chamber in Fig.~\ref{FIG1}(b) could exhibit a transient counterclockwise rotation as $\boldsymbol{G}\times\boldsymbol{v}=-\boldsymbol{F}_{\text{s}}$ when $F_{\text{s}}\neq 0$. We will provide simulation evidence for this phenomenon later.}

In Figs.~\ref{FIG1}(c) and~\ref{FIG1}(d), we show the simulated system where many skyrmions can interact with \emph{an OAG} separating two chambers.
These skyrmions can also interact with nearby skyrmions inside the chamber. Namely, both the skyrmion-wall and skyrmion-skyrmion interactions can play roles in the simulated system.
The system is a rectangle ferromagnetic monolayer with interface-induced DM interaction and perpendicular magnetic anisotropy (PMA), which has typical magnetic parameters~\cite{Tomasello_SREP2014,Xichao_PRB2022B,Zhang_Laminar2023,Zhang_NL2023}: the saturation magnetization $M_{\text{S}}=580$ kA m$^{-1}$, exchange constant $A=15$ pJ m$^{-1}$, PMA constant $K=0.8$ MJ m$^{-3}$, and DM interaction parameter $D=3$ mJ m$^{-2}$. The length, width, and thickness of the system are equal to $600$ nm, $200$ nm, and $1$ nm, respectively.
By locally enhancing PMA, we could fabricate barrier walls on the ferromagnetic layer~\cite{Juge_NL2021,Ohara_NL2021,Zhang_CP2021,Zhang_NL2023}, which form a confined geometry where two chambers with default PMA are separated by a wall with enhanced PMA of $10K$. The barrier wall width (in the $x$ dimension) is fixed at $28$ nm.
Skyrmions cannot penetrate the barrier wall, however, a gate on the barrier wall with the opening width of $W$ may allow skyrmions to pass through [Fig.~\ref{FIG1}(a)].
The gate placed near the upper chamber wall forms \emph{an OAG} [Figs.~\ref{FIG1}(c) and~\ref{FIG1}(d)], while the gate placed at the chamber center forms \emph{a centered symmetric gate (CSG)} [Figs.~\ref{FIG1}(e) and~\ref{FIG1}(f)].

As discussed in Figs.~\ref{FIG1}(a) and~\ref{FIG1}(b), the skyrmion with $Q=-1$ is in principle easier to pass through the given \emph{OAG} [Fig.~\ref{FIG1}(c)] from the left chamber provided that the opening width $W$ is appropriate. It is also expected that the possibility to pass through \emph{a CSG} [Fig.~\ref{FIG1}(e)] should be identical for the left-to-right ($\text{L}\rightarrow\text{R}$) and right-to-left ($\text{R}\rightarrow\text{L}$) cases.
We therefore first \emph{qualitatively} analyze possible diffusive behaviors of many skyrmions in the \emph{OAG and CSG} systems from the viewpoint of thermodynamics~\cite{Callen_1985}.

We consider $20$ skyrmions in the left or right chamber at $t=0$, and assume a simplified system with a fixed temperature and no thermal annihilation of skyrmions. We also assume that the interactions between skyrmions and the gate are always effective and do not depend on other factors, such as the time-dependent skyrmion density. Thus, the diffusion rate of skyrmions is treated as a time-independent parameter.
As skyrmions with $Q=-1$ are easier to pass through the \emph{OAG} from the left, the diffusion rate $k_{\text{L}\rightarrow\text{R}}$ should be larger than $k_{\text{R}\rightarrow\text{L}}$, where the subscript ``$\text{L}\rightarrow\text{R}$'' denotes that skyrmions diffuse from the left to the right chamber, and ``$\text{R}\rightarrow\text{L}$'' denotes that skyrmions diffuse from the right to the left chamber. We assume $k_{\text{L}\rightarrow\text{R}}=0.23$ and $k_{\text{R}\rightarrow\text{L}}=0.1$.
\emph{These values of $k_{\text{L}\rightarrow\text{R}}$ and $k_{\text{R}\rightarrow\text{L}}$ are chosen to illustrate representative asymmetric diffusive behavior that can be compared qualitatively with the simulation results.}

Therefore, the equilibrium probability distribution of skyrmions in the left chamber equals $P(n)=\Omega(n)(p_{\text{L}})^{n}(p_{\text{R}})^{N-n}$, where $p_{\text{L}}=k_{\text{R}\rightarrow\text{L}}/(k_{\text{L}\rightarrow\text{R}}+k_{\text{R}\rightarrow\text{L}})$ and $p_{\text{R}}=k_{\text{L}\rightarrow\text{R}}/(k_{\text{L}\rightarrow\text{R}}+k_{\text{R}\rightarrow\text{L}})=1-p_{\text{L}}$ are the probabilities of a skyrmion being in the left and right chamber, respectively.
$\Omega(n)=N!/[n!(N-n)!]$ represents the number of microstates when the number of skyrmions in the left chamber $N_{\text{L}}=n$. $N_{\text{T}}=20$ is the total number of skyrmions in the system.
In Fig.~\ref{FIG1}(g), it shows that the asymmetric diffusion $k_{\text{L}\rightarrow\text{R}}>k_{\text{R}\rightarrow\text{L}}$ leads to different equilibrium distributions of skyrmions, where more skyrmions can be found in the right chamber at equilibrium. However, for symmetric diffusion (for example, $k_{\text{L}\rightarrow\text{R}}=k_{\text{R}\rightarrow\text{L}}=0.165$), the equilibrium distributions of skyrmions are identical in the left and right chambers [Fig.~\ref{FIG1}(h)].

In Fig.~\ref{FIG1}(i), we show the theoretical diffusion if a total number of $N_{\text{L}}(0)=N_{\text{T}}$ skyrmions are placed in the left chamber initially [i.e., $N_{\text{R}}(0)=0$].
The diffusion rate equations for the number of skyrmions in the left and right chambers read $N_{\text{L}}(t)=p_{\text{L}}N_{\text{T}}+(N_{\text{T}}-p_{\text{L}}N_{\text{T}})e^{-(k_{\text{L}\rightarrow\text{R}}+k_{\text{R}\rightarrow\text{L}})t}$ and $N_{\text{R}}(t)=\emph{N_{\text{T}}}-N_{\text{L}}(t)$, respectively.
The case with $N_{\text{L}}(0)=0$ and $N_{\text{R}}(0)=N_{\text{T}}$ is given in Fig.~\ref{FIG1}(j).
On the other hand, the solutions of the rate equations for skyrmions interacting with \emph{a CSG} ($k_{\text{L}\rightarrow\text{R}}=k_{\text{R}\rightarrow\text{L}}$) are given in Figs.~\ref{FIG1}(k) and~\ref{FIG1}(l).

For ease of observation, we focus on the initially empty chamber and count the skyrmions diffused to it within a fixed period of time.
We find asymmetric diffusion for skyrmions interacting with \emph{an OAG}, where more skyrmions can escape from the left chamber [Fig.~\ref{FIG1}(m)] within the same time.
However, for skyrmions interacting with \emph{a CSG}, the diffusion is symmetric and one can find the same amount of escaped skyrmions within the same time [Fig.~\ref{FIG1}(n)].

\emph{It should be noted that in this work the initial condition $N_{\text{R}}(0)=0$ or $N_{\text{L}}(0)=0$ is adopted to isolate the directional asymmetry in the clearest possible manner, without additional complications caused by an initial skyrmion imbalance between the two connected chambers. This setting also makes the diffusion asymmetry easier to observe in the simulation experiments shown below.}

%-%-%-%-%-%-%-%-%-%-%-%-%-%-%-%-%-%-%-%-%-%-%-%-%-%-%-%-%-%-%
%\section{Computational Experiments}
%\label{se:Simulation}
%-%-%-%-%-%-%-%-%-%-%-%-%-%-%-%-%-%-%-%-%-%-%-%-%-%-%-%-%-%-%

%-%-%-%-%-%-%-%-%-%-%-%-%-%-%-%-%-%-%-%-%-%-%-%-%-%-%-%-%-%-%
\vbox{}
\noindent
\textbf{Computational experiments}
%-%-%-%-%-%-%-%-%-%-%-%-%-%-%-%-%-%-%-%-%-%-%-%-%-%-%-%-%-%-%

\noindent
To examine the theoretical predictions, we carried out computational experiments by simulating the spin dynamics using the MuMax$^3$ simulator~\cite{MuMax}.
The spin dynamics at finite temperature is governed by the stochastic Landau-Lifshitz-Gilbert equation (see Methods).

In Figs.~\ref{FIG2}(a) and~\ref{FIG2}(b), we show simulated diffusion of skyrmions with $Q=-1$ interacting with \emph{an OAG} (see \blue{Supplementary Videos 1} and \blue{2}). We assume $W=36$ nm and $T=150$ K with a thermal random seed of $S=304$. At $t=0$, $20$ skyrmions with $Q=-1$ are relaxed in either the left or right chamber.
For the case with $N_{\text{L}}(0)=20$ and $N_{\text{R}}(0)=0$, we find $N_{\text{R}}(500~\text{ns})=6$ and $N_{\text{R}}(1000~\text{ns})=9$.
For the case with $N_{\text{L}}(0)=0$ and $N_{\text{R}}(0)=20$, we find $N_{\text{L}}(500~\text{ns})=2$ and $N_{\text{L}}(1000~\text{ns})=3$.
These simulation outcomes indicate the skyrmion diffusion in the \emph{OAG} system is asymmetric.
In contrast, the observation of the \emph{CSG} system indicates a nearly symmetric diffusion (see \blue{Supplementary Videos 3} and \blue{4}) at $t=500$-$1000$ ns [Figs.~\ref{FIG2}(c) and~\ref{FIG2}(d)].

In Figs.~\ref{FIG2}(e) and~\ref{FIG2}(f), we show $N_{\text{L}}(t)$ and $N_{\text{R}}(t)$ corresponding to the systems given in Figs.~\ref{FIG2}(a) and~\ref{FIG2}(b) for $1000$-ns-long simulations with the same parameters, respectively.
$N_{\text{L} (\text{L}\rightarrow\text{R})}$ and $N_{\text{R} (\text{L}\rightarrow\text{R})}$ represent the number of skyrmions in the left and right chambers, respectively, during the left-to-right ($\text{L}\rightarrow\text{R}$) diffusion process, where skyrmions initially placed in the left chamber diffuse into the right chamber [Figs.~\ref{FIG2}(a) and~\ref{FIG2}(c)].
Conversely, $N_{\text{L} (\text{R}\rightarrow\text{L})}$ and $N_{\text{R} (\text{R}\rightarrow\text{L})}$ indicate the number of skyrmions in each chamber during the right-to-left ($\text{R}\rightarrow\text{L}$) diffusion process [Figs.~\ref{FIG2}(b) and~\ref{FIG2}(d)].
By comparing $N_{\text{R} (\text{L}\rightarrow\text{R})}$ and $N_{\text{L} (\text{R}\rightarrow\text{L})}$, the diffusion asymmetry in the \emph{OAG} system is confirmed [Fig.~\ref{FIG2}(g)], which is \emph{qualitatively} in line with the theoretical trend in Fig.~\ref{FIG1}(m).
For the \emph{CSG} system, $N_{\text{R} (\text{L}\rightarrow\text{R})}$ is very close to $N_{\text{L} (\text{R}\rightarrow\text{L})}$ [Fig.~\ref{FIG2}(j)], which agrees with the theoretical trend [Fig.~\ref{FIG1}(n)].

\emph{It should be noted that the total number of skyrmions in simulations is decreasing with time due to thermal annihilation [Fig.~\ref{FIG3}], which could reduce the skyrmion density and affect the observation outcome.}
Namely, the diffusion rate depends on the skyrmion density that affects the effectiveness of skyrmion-wall and skyrmion-skyrmion interactions, therefore, it is expected that the diffusion rates $k_{\text{L}\rightarrow\text{R}}$ and $k_{\text{R}\rightarrow\text{L}}$ in simulations are time-dependent and may approach the same value even if they start from $k_{\text{L}\rightarrow\text{R}}>k_{\text{R}\rightarrow\text{L}}$ at $t=0$.

\emph{Also, we note that even the initial skyrmion density given in the simulated systems at $t=0$ can significantly affect the diffusion rate from the outset [Fig.~\ref{FIG4}], which is a distinct property inherent to interacting skyrmion systems.}
\emph{Namely, the diffusion of skyrmions through the \emph{OAG} or \emph{CSG} also depends on the initial density of skyrmions inside the chamber.
To investigate this, we carried out simulations with different total number of skyrmions based on the systems given in Fig.~\ref{FIG2}.
In Fig.~\ref{FIG4}, it can be seen that the diffusion rate of skyrmions increases with the initial total number of skyrmions placed in the left or right chamber for both the \emph{OAG} [Figs.~\ref{FIG4}(a)-\ref{FIG4}(d)] and \emph{CSG} [Figs.~\ref{FIG4}(e)-\ref{FIG4}(h)] systems.
For example, when $4$ skyrmions are placed in the system at $t=0$, no skyrmion is diffused into the opposite chamber in the \emph{OAG} system [Fig.~\ref{FIG4}(i)] during $t=0$-$1000$ ns. For the \emph{CSG} system [Fig.~\ref{FIG4}(l)], only one skyrmion is diffused from the right chamber to the left during $t=0$-$1000$ ns.
When $20$ skyrmions are placed in the system at $t=0$, the asymmetric and symmetric diffusion are found in the \emph{OAG} [Fig.~\ref{FIG4}(j)] and \emph{CSG} [Fig.~\ref{FIG4}(m)] systems, respectively.
When $28$ skyrmions are placed in the system at $t=0$, the diffusion symmetry still depends on the gate geometry but the diffusion strengths in both the \emph{OAG} [Fig.~\ref{FIG4}(k)] and \emph{CSG} [Fig.~\ref{FIG4}(n)] systems are more significant during the initial stage of the diffusion (i.e., $t=0$-$200$ ns).}

Furthermore, since skyrmions have a finite size and interact through short-range repulsive forces, an excessively high skyrmion density could lead to strong skyrmion-skyrmion repulsion. This repulsion may cause skyrmions to be expelled from the chamber without effective interaction with the \emph{OAG} geometry.
In our model, $20$ skyrmions in either the left or right chamber results in a moderate initial density that ensures $k_{\text{L}\rightarrow\text{R}}>k_{\text{R}\rightarrow\text{L}}$ in the \emph{OAG} system.

We also performed $100$ repetitions of the simulations demonstrated in Fig.~\ref{FIG2} with the same parameters but different thermal random seeds $S$.
In Fig.~\ref{FIG2}(k), we show the summation of the number of skyrmions observed at $t=500$ ns in the initially empty right chamber $\sum_{S} N_{\text{R} (\text{L}\rightarrow\text{R})}$ over that in the initially empty left chamber $\sum_{S} N_{\text{L} (\text{R}\rightarrow\text{L})}$.
For the \emph{OAG} system, we find $\sum_{S} N_{\text{R} (\text{L}\rightarrow\text{R})} / \sum_{S} N_{\text{L} (\text{R}\rightarrow\text{L})}>1$ for $100$ repetitions, justifying the asymmetric diffusion.
For the \emph{CSG} system, we find $\sum_{S} N_{\text{R} (\text{L}\rightarrow\text{R})} / \sum_{S} N_{\text{L} (\text{R}\rightarrow\text{L})}$ generally approaches $1$, justifying the symmetric diffusion.

We also compare $\sum_{S} N_{\text{R} (\text{L}\rightarrow\text{R})}^{\text{OAG}}$ of the \emph{OAG} system with $\sum_{S} N_{\text{R} (\text{L}\rightarrow\text{R})}^{\text{CSG}}$ of the \emph{CSG} system [Fig.~\ref{FIG2}(l)].
$\sum_{S} N_{\text{R} (\text{L}\rightarrow\text{R})}^{\text{OAG}}/\sum_{S} N_{\text{R} (\text{L}\rightarrow\text{R})}^{\text{CSG}}$ is smaller than $1$, which means that the skyrmions are easier to pass through the \emph{CSG} compared to the \emph{OAG} system.
This reason is that the \emph{CSG} is positioned at the system center, whereas the \emph{OAG} is located near the upper border. Consequently, skyrmions are more likely to interact with the \emph{CSG} than with the \emph{OAG}.
\emph{Such a trend is pronounced when more repetitions are performed [Fig.~\ref{FIG5}].}
On the other hand, the value of $\sum_{S} N_{\text{R} (\text{L}\rightarrow\text{R})}^{\text{OAG}}/\sum_{S} N_{\text{R} (\text{L}\rightarrow\text{R})}^{\text{CSG}}$ is obviously larger than that of $\sum_{S} N_{\text{L} (\text{R}\rightarrow\text{L})}^{\text{OAG}}/\sum_{S} N_{\text{L} (\text{R}\rightarrow\text{L})}^{\text{CSG}}$, which is a result of the diffusion asymmetry in the \emph{OAG} system [$\sum_{S} N_{\text{R} (\text{L}\rightarrow\text{R})}^{\text{OAG}}>\sum_{S} N_{\text{L} (\text{R}\rightarrow\text{L})}^{\text{OAG}}$] as well as the symmetric diffusion in the \emph{CSG} system [$\sum_{S} N_{\text{R} (\text{L}\rightarrow\text{R})}^{\text{CSG}}\sim\sum_{S} N_{\text{L} (\text{R}\rightarrow\text{L})}^{\text{CSG}}$].

In Fig.~\ref{FIG6}, we highlight the simulated spin dynamics that leads to the asymmetric diffusion in the \emph{OAG} system [Fig.~\ref{FIG2}(a)].
At $t=410$ ns, a diffusive skyrmion in the left chamber and near the \emph{OAG} slowly approaches the upper chamber wall [Fig.~\ref{FIG6}(a)], and it then accelerates and moves along the upper chamber wall toward the right chamber, which agrees with the Thiele model analysis [Fig.~\ref{FIG1}(a)]. The skyrmion slightly shrinks when it moves along the wall and successfully diffuse into the right chamber at $t=425$ ns.
In Fig.~\ref{FIG6}(b), it can be seen that a skyrmion in the right chamber first approaches the gate wall and moves toward the upper chamber wall. The skyrmion turns right when it detaches from the gate wall but touches the upper chamber wall near the gate, as predicted by the Thiele model [Fig.~\ref{FIG1}(b)]. As a result, the skyrmion diffuses away from the gate and remains in the right chamber.
If the opening width of the gate is much wider than the skyrmion size, the skyrmion may pass through the gate without effective interaction with the upper chamber wall.

\emph{In Fig.~\ref{FIG7}, we show that the opening width $W$ of the gate could affect the statistical outcome as the skyrmion-wall interaction delicately depends on the ratio between the skyrmion size and $W$.}
\emph{An OAG with $W$ much narrower than the skyrmion size does not allow the skyrmion to pass through, while an OAG with $W$ much wider than the skyrmion size allow all skyrmions to pass through without effective skyrmion-gate interaction. Only a reasonable opening width $W$ close to the skyrmion size can lead to the asymmetric diffusion of skyrmions due to effective interactions among skyrmions and gate/chamber walls.
It can be seen from Fig.~\ref{FIG7} that skyrmions are able to diffuse into the opposite chamber when $W$ is larger than $30$ nm and the asymmetric diffusion could be observed when $W=30\sim 32$ nm and $W=36\sim 38$ nm.
On the other hand, Fig.~\ref{FIG7} shows that when $W=0$ nm, the diffusion between the left and right chambers cannot happen, all skyrmions will remain confined within the chamber in which they were initially placed, and some skyrmions will be thermally annihilated during their diffusion. When the system has no gate at all (i.e., $W=192$ nm), all skyrmions are free to diffuse between the left and right chambers, and basically there is no diffusion asymmetry as observed at $t=500$ ns.
An interesting point is that the total number of skyrmions in the system without a gate (i.e., $W=192$ nm) at $t=500$ ns is larger than that in the system with $W=0$ nm, which means that diffusive skyrmions are more likely to annihilate in a smaller confined space, where skyrmions can experience stronger interactions with neighboring skyrmions and walls.
The compression between skyrmions is a known factor that leads to their annihilation~\cite{Zhang_PRB2022_Compression}. Effective thermal diffusion of skyrmions between two connecting chambers can significantly reduce the inter-skyrmion compression within a short period, thereby increasing the number of skyrmions that can survive in thermal systems.}

In Fig.~\ref{FIG6}(c), we also show that two diffusive skyrmions can demonstrate rotational behavior when they meet each other inside the chamber. \emph{Such a counterclockwise rotation can be explained by the Thiele model considering the skyrmion-skyrmion interaction [Fig.~\ref{FIG1}(b)], and is a unique dynamic phenomenon that can be found only in systems with interacting skyrmions.} It suggests that even the repulsive interaction between two diffusive skyrmions may bond them into a transient binary system.

%-%-%-%-%-%-%-%-%-%-%-%-%-%-%-%-%-%-%-%-%-%-%-%-%-%-%-%-%-%-%
%\section{Conclusion}
%\label{se:Conclusion}
%-%-%-%-%-%-%-%-%-%-%-%-%-%-%-%-%-%-%-%-%-%-%-%-%-%-%-%-%-%-%

%-%-%-%-%-%-%-%-%-%-%-%-%-%-%-%-%-%-%-%-%-%-%-%-%-%-%-%-%-%-%
\vbox{}
\noindent
\textbf{Discussion}
%-%-%-%-%-%-%-%-%-%-%-%-%-%-%-%-%-%-%-%-%-%-%-%-%-%-%-%-%-%-%

\noindent
We have evidenced the asymmetric diffusion of many \emph{repulsive} skyrmions interacting with \emph{an off-center asymmetric} gate. Such a highly nontrivial dynamic phenomenon is a result of the Brownian gyromotion in structured environment, reflecting the topological nature of skyrmions as well as their mutual interactions.
\emph{Namely, the asymmetric diffusion does not originate from a geometrically asymmetric gate by itself, but from the interplay between the asymmetric confinement geometry and the dynamics of skyrmions.}
\emph{First, thermal fluctuations bring the skyrmion into contact with the relevant chamber wall or gate wall; once the skyrmion approaches a wall, the wall exerts a repulsive force normal to the boundary; the gyrotropic term then converts this repulsive force into a transverse drift along the boundary. Because the OAG selectively exposes the skyrmion to different wall segments depending on the chamber from which it approaches, the resulting wall-guided trajectories generally favor one crossing direction over the other.}

\blue{Moreover, we point out that although we considered a fixed small damping in order to see more active Brownian behavior of skyrmions in our computational simulations, the damping parameter cannot be zero in principle, and therefore the nonzero dissipative force $\left|\alpha\boldsymbol{\D}\cdot\boldsymbol{v}\right|$ could delicately modify the velocity field of skyrmions near the OAG, as illustrated in Fig.~\ref{FIG8}.
Namely, the skyrmion-wall interaction is actually described by $\boldsymbol{G}\times\boldsymbol{v}-\alpha\boldsymbol{\D}\cdot\boldsymbol{v}=-\boldsymbol{F}_{\text{w}}$, where $\left|\boldsymbol{G}\times\boldsymbol{v}\right|>\left|\alpha\boldsymbol{\D}\cdot\boldsymbol{v}\right|\neq 0$.
As $\boldsymbol{F}_{\text{w}}$ is always perpendicular to the wall, the nonzero dissipative term will result in the fact that the skyrmion trajectory in response to $\boldsymbol{F}_{\text{w}}$ is not perfectly parallel to the wall. Namely, the trajectory is actually slightly tilted away from the wall.
Consequently, the velocity field around the OAG indicates that skyrmions approaching from the left guided by the upper chamber wall appear to have a natural path into the gate, while skyrmions approaching from the right are naturally funneled away from the constriction by the tilted velocity field [Fig.~\ref{FIG8}(a)].
Such a physical picture suggests that the diffusion asymmetry arises not only from the wall-guided gyrotropic motion, but from the near-OAG-region velocity field produced by the combined gyrotropic and dissipative response to the repulsive skyrmion-wall interaction, where the nonzero damping plays an essential role.
Figure~\ref{FIG8}(b) illustrates the case where the damping is set to zero and the dissipative term is totally removed (i.e., $\left|\alpha\boldsymbol{\D}\cdot\boldsymbol{v}\right|=0$). It shows that the asymmetric diffusion is more difficult to happen as skyrmions can equally easily pass through the gate from both sides (guided by the chamber wall or gate wall).
As the nonzero damping plays a crucial role in thermally driven skyrmion dynamics~\cite{Raj_N2024,Gong_APL2022} by affecting several factors in a complicated way, including the diffusion coefficient, the skyrmion lifetime, and the dissipative term, future studies could further explore its potential effects on diffusion asymmetry.}

The diffusion asymmetry also depends on other factors, such as the gate \emph{symmetry}, gate opening width, and skyrmion density.
\emph{We note that the rate equation is intended as a qualitative description, whereas in the simulated system the effective diffusion rates are time-dependent because the skyrmion density evolves due to diffusion and thermal annihilation. The comparison between the rate-equation solutions and the simulation results should therefore be understood as qualitative in trend.}
We also note that the gate cannot be modified once fabricated but skyrmion density and size are subject to thermal effects and external magnetic fields, for example. The size of a typical skyrmion or skyrmion bubble could be a few hundred nanometers in experimental samples, for example, in CoFeB thin film~\cite{Wanjun_NPHYS2017}, the skyrmion bubble size is about $800$-$1000$ nm. Therefore, it will not be difficult to fabricate a gate with an opening width slightly larger than the skyrmion size. The size of a skyrmion can also be flexibly adjusted by applying an external magnetic field perpendicular to the sample in experiments.
Because the diffusion rate in real system is dependent of the skyrmion density and is varying with time, the simulated system may not reach a sustained nonequilibrium steady state but could create time-dependent nonequilibrium phenomena before it reaches thermal equilibrium driven by the increasing entropy of the entire system.
The effects of \emph{repulsive} skyrmion-skyrmion interaction and skyrmion density on their diffusion is an essential feature that can only be found in interacting skyrmion systems.
In our simulated systems, we also observed that diffusive skyrmions could demonstrate emergent rotational bond dynamics due to repulsive skyrmion-skyrmion interaction.

\emph{Future studies may also investigate the diffusion asymmetry of Bloch-type skyrmions in structured environment, however, it should be noted that the sign of the gyrotropic response is determined by the topological charge, not by the helicity. Therefore, the diffusion asymmetry could be largely independent of skyrmion helicity, and may persist even in systems with mixed helicity structures.}
\emph{In addition, some other systems may show attractive skyrmion-skyrmion and skyrmion-boundary interactions, for example, in certain bulk chiral magnets. In such cases the detailed behavior could be altered significantly due to attractive skyrmion-edge and skyrmion-skyrmion interactions. For example, skyrmions may form clusters and thus, their interaction with the gate as well as their general diffusive behavior could be different to that of repulsive skyrmions. The sign, strength, and range of the boundary interaction will affect both the wall-guided drift and the optimal gate width. An attractive edge may weaken, modify, or even reverse the guidance mechanism reported in this work, depending on the detailed potential landscape. These cases would be highly interesting to investigate in future studies.}

This work paves the way for the study of complex diffusion behavior of chiral magnetic matter, which may be useful for building nonconventional computing devices based on diffusion asymmetry.
\emph{For example, the diffusion asymmetry in physical particle systems can be used for AI, especially for physical reservoir computing, neuromorphic computing, stochastic and probabilistic computing devices.}
Our results should also be relevant to other systems where chiral \emph{or asymmetric} diffusion can occur~\cite{Zuiden_PNAS2016,Banerjee_NC2017,Fruchart_NATURE2021,Fruchart_ARCMP2023}, such as various types of chiral active matter systems, systems with chiral nonreciprocal interactions, systems with odd viscosity, and certain charged systems in magnetic fields.

%-%-%-%-%-%-%-%-%-%-%-%-%-%-%-%-%-%-%-%-%-%-%-%-%-%-%-%-%-%-%
\vbox{}
\noindent
\textbf{Methods}
%-%-%-%-%-%-%-%-%-%-%-%-%-%-%-%-%-%-%-%-%-%-%-%-%-%-%-%-%-%-%

%-%-%-%-%-%-%-%-%-%-%-%-%-%-%-%-%-%-%-%-%-%-%-%-%-%-%-%-%-%-%
\noindent
\textbf{Computational simulations}
%-%-%-%-%-%-%-%-%-%-%-%-%-%-%-%-%-%-%-%-%-%-%-%-%-%-%-%-%-%-%

\noindent
In this work, we carried out computational experiments by simulating the spin dynamics of the entire system.
We consider spin dynamics at finite temperature governed by the stochastic Landau-Lifshitz-Gilbert equation $\partial_{t}\boldsymbol{m}=-\gamma_{0}\boldsymbol{m}\times(\boldsymbol{h}_{\text{eff}}+\boldsymbol{h}_{\text{f}})+\alpha(\boldsymbol{m}\times\partial_{t}\boldsymbol{m})$,
where $\boldsymbol{m}$ is the reduced magnetization,
$\gamma_{0}=2.211\times 10^{5}$ m A$^{-1}$ s$^{-1}$ is the absolute gyromagnetic ratio,
$\alpha=0.1$ is the Gilbert damping parameter,
$\boldsymbol{h}_{\rm{eff}}=-\frac{1}{\mu_{0}M_{\text{S}}}\cdot\frac{\delta\varepsilon}{\delta\boldsymbol{m}}$ is the effective field.
$\mu_{0}$ and $\varepsilon$ denote the vacuum permeability constant and average energy density, respectively.
$\boldsymbol{h}_{\text{f}}$ is a thermal fluctuating field satisfying
$<h_{i}(\boldsymbol{x},t)>=0$ and
$<h_{i}(\boldsymbol{x},t)h_{j}(\boldsymbol{x}',t')>=\frac{2\alpha k_{\text{B}}T}{M_\text{S}\gamma_0\mu_0 V}\delta_{ij}\delta(\boldsymbol{x}-\boldsymbol{x}')\delta(t-t')$,
where $i$ and $j$ are Cartesian components, $k_{\text{B}}$ is the Boltzmann constant, $T$ is the temperature, and $V$ is the volume of a single mesh cell.
$\delta_{ij}$ and $\delta(\dots)$ denote the Kronecker and Dirac delta symbols, respectively.
The energy terms include the ferromagnetic exchange, interface-induced DM interaction, PMA, and demagnetization. Hence, the average energy density is expressed as
$\varepsilon=A\left(\nabla\boldsymbol{m}\right)^{2}+D\left[m_{z}\left(\boldsymbol{m}\cdot\nabla\right)-\left(\nabla\cdot\boldsymbol{m}\right)m_{z}\right]-K(\boldsymbol{n}\cdot\boldsymbol{m})^2-\frac{M_{\text{S}}}{2}(\boldsymbol{m}\cdot\boldsymbol{B}_{\text{d}})$
with $\boldsymbol{B}_{\text{d}}$ being the demagnetization field. $\boldsymbol{n}$ is the unit surface normal vector. $m_z$ is the out-of-plane component of $\boldsymbol{m}$.
The mesh size is set to $2$ $\times$ $2$ $\times$ $1$ nm$^3$ and a fixed finite-temperature integration time step of $10$ fs is applied.

%-%-%-%-%-%-%-%-%-%-%-%-%-%-%-%-%-%-%-%-%-%-%-%-%-%-%-%-%-%-%
\vbox{}
\noindent
\textbf{Data availability}
%-%-%-%-%-%-%-%-%-%-%-%-%-%-%-%-%-%-%-%-%-%-%-%-%-%-%-%-%-%-%

\noindent
The authors declare that the main data supporting the findings of this study are available within the article and its Supplementary Information files. Extra data are available from the corresponding authors upon reasonable request.

%-%-%-%-%-%-%-%-%-%-%-%-%-%-%-%-%-%-%-%-%-%-%-%-%-%-%-%-%-%-%
\vbox{}
\noindent
\textbf{Code availability}
%-%-%-%-%-%-%-%-%-%-%-%-%-%-%-%-%-%-%-%-%-%-%-%-%-%-%-%-%-%-%

\noindent
The micromagnetic simulator MUMAX3 used in this work is publicly accessible at
http://mumax.github.io/index.html.

%-%-%-%-%-%-%-%-%-%-%-%-%-%-%-%-%-%-%-%-%-%-%-%-%-%-%-%-%-%-%

%-%-%-%-%-%-%-%-%-%-%-%-%-%-%-%-%-%-%-%-%-%-%-%-%-%-%-%-%-%-%

%-%-%-%-%-%-%-%-%-%-%-%-%-%-%-%-%-%-%-%-%-%-%-%-%-%-%-%-%-%-%
\vbox{}
\noindent
\textbf{Acknowledgements}
%-%-%-%-%-%-%-%-%-%-%-%-%-%-%-%-%-%-%-%-%-%-%-%-%-%-%-%-%-%-%

\noindent
X.Z. and M.M. acknowledge support by the CREST, the Japan Science and Technology Agency (Grant No. JPMJCR20T1). X.Z. also acknowledges support by the Grants-in-Aid for Scientific Research from JSPS KAKENHI (Grant No. JP25K17939 and No. JP20F20363). M.M. also acknowledges support by the Grants-in-Aid for Scientific Research from JSPS KAKENHI (Grants No. JP25H00611, No. JP24H02231, No. JP23H04522, and No. JP20H00337) and the Waseda University Grant for Special Research Projects (Grant No. 2025C-133).
C.R. and C.J.O.R. acknowledge the support by the U.S. Department of Energy through the Los Alamos National Laboratory. Los Alamos National Laboratory is operated by Triad National Security, LLC, for the National Nuclear Security Administration of the U.S. Department of Energy (Contract No. 892333218NCA000001).
\emph{Q.S. acknowledges funding support from RGC GRF (16309924), NSFC/RGC JRS (N\_HKUST675/24), and MOST Key R\&D Program (MOST22SC01).}
R.Z. acknowledges the support by the State Key Laboratory of Displays and Opto-Electronics (Project Reference: ITC-PSKL12EG02).
\emph{R.Z. also acknowledges financial support from Research Grants Council of Hong Kong via Grant No. 16306924.}
Y.Z. acknowledges support by the Shenzhen Fundamental Research Fund (Grant No. JCYJ20210324120213037), the Guangdong Basic and Applied Basic Research Foundation (Grant No. 2021B1515120047), the Shenzhen Peacock Group Plan (Grant No. KQTD20180413181702403), and the National Natural Science Foundation of China (Grant No. 12374123), Guangdong Basic Research Center of Excellence for Aggregate Science, and the 2023 SZSTI stable support scheme.
Y.X. acknowledges support by the National Key Research and Development Program of China (Grants No. 2024YFA1408801) and the National Natural Science Foundation of China (Grant No. 62427901).

%-%-%-%-%-%-%-%-%-%-%-%-%-%-%-%-%-%-%-%-%-%-%-%-%-%-%-%-%-%-%
\vbox{}
\noindent
\textbf{Author contributions}
%-%-%-%-%-%-%-%-%-%-%-%-%-%-%-%-%-%-%-%-%-%-%-%-%-%-%-%-%-%-%

\noindent
X.Z. and M.M. co-designed the project. \emph{X.Z., Q.S., R.Z., and M.M. coordinated the interdisciplinary collaboration.} X.Z. performed the simulations. X.Z., C.R., C.J.O.R., Q.S., R.Z., Y.Z., Y.X., and M.M. analyzed the data. X.Z. drafted the paper with input from C.R., C.J.O.R., Q.S., R.Z., Y.Z., Y.X., and M.M. All authors discussed the results and contents of the paper.

%-%-%-%-%-%-%-%-%-%-%-%-%-%-%-%-%-%-%-%-%-%-%-%-%-%-%-%-%-%-%
\vbox{}
\noindent
\textbf{Competing interests}
%-%-%-%-%-%-%-%-%-%-%-%-%-%-%-%-%-%-%-%-%-%-%-%-%-%-%-%-%-%-%

\noindent
\emph{Q.S. is Associate Editor of npj Spintronics. Q.S. was not involved in the journal's review of, or decisions related to, this manuscript. The remaining authors declare no competing financial or non-financial interests.}

%%%%%%%%%%%%%%%%%%%%%%%%%%%%%%%%%%%%%%%%%%%%%%%%%%%%%%%%%%%%
\end{document}